# Prediction of high-$T_c$ superconductivity in two corrugated graphene sheets with intercalated CeH$_9$ molecules

M. A. Rastkhadiv[1] M. Pazoki[2,3]

[1] Estahban Higher Education Center, Shiraz University, Estahban, Iran
[2] Department of Chemistry, Ångström Laboratory,
Uppsala University, Box 538, 75121 Uppsala, Sweden
[3] Institute for Photovoltaics, Stuttgart University, Stuttgart 70569, Germany

Recent discoveries involving high-temperature superconductivity in H$_3$S and LaH$_{10}$ have sparked a renewed interest in exploring the potential for superconductivity within hydrides. These superconductors require extremely high-pressure condition ($\sim 100\ GPa$), rendering them virtually impractical for industrial applications. In this study, we verify the occurrence of a low pressure superconductivity phase transition in a system containing two graphene layers with sine form corrugations where CeH$_9$ doped molecules are intercalated between the layers. The lowest-order constrained variational method is applied to calculate the thermodynamic and electrical properties of the valence electrons. We examine 9900 different distributions of CeH$_9$ molecules separately for finding a second-order phase transition with maximized critical temperature. The novelty of the present work is the prediction of a superconductivity transition at $T_c = 198.61\ K$ for a specific distribution of CeH$_9$ molecules with applying no external pressure on the exterior surfaces of the graphene sheets. Notably, this critical temperature is approximately 65 $K$ higher than that observed in cuprate materials (HgBa$_2$Ca$_2$Cu$_3$O$_{8+\delta}$), which are known for their high $T_c$ values at room pressure. It is interesting that in this particular case, the distribution periodicity of CeH$_9$ molecules bears the closest resemblance to the periodicity of the graphene corrugations among all 9900 examined cases. Computing the energy gap of the valence electrons reveals that this critical behavior corresponds to an unconventional superconductivity phase transition exhibiting a high critical current density on the order of $\sim 10^7\ A/cm^2$.

## I. INTRODUCTION

After the discovery of the first superconductor, condensed matter physicists have been diligently working to identify materials with higher temperatures for achieving superconductivity. Recent advancements in this pursuit include the discovery of high-temperature superconductivity in superhydrides, which exhibit critical temperatures ($T_c$s) reaching up to 250 $K$ under extreme high-pressures on order of $\sim 170\ GPa$[1]. These remarkable findings mark a new era characterized by significant technological achievements. The extreme high-pressure condition remains a substantial challenge in applying these types of superconductors in everyday applications. As a result, cuprate materials (HgBa$_2$Ca$_2$Cu$_3$O$_{8+\delta}$) with $T_c = 133.5\ K$ are currently regarded as the highest temperature superconductors at room pressure [2, 3]. Recent investigations into enhancing the critical temperature of superconductivity through the application of superhydrides have suggested two essential conditions for achieving $T_c > 133.5\ K$: extreme high-pressure [4–8], and hydrogen-rich property [9, 10]. At extreme high-pressure conditions, the primary factor that leads to superconductivity in superhydrides and increases the formation of Cooper pairs is the high electron-phonon coupling [9, 10].

Theoretical approaches such as density functional theory (DFT), built upon Migdal-Eliashberg theory [11], have exerted a significant influence in predicting the emergence of high-temperature superconductivity within clathrate-like structures. In these compounds, a host atom (Y[12, 13], Ca [14], and La [12, 15, 16]) occupies the central position within H$_n$ cages and donates its electrons to stabilize the clathrate structure. These cages contain numerous hydrogen atoms, H$_{24}$ for CaH$_6$ [14] and YH$_6$ [13], H$_{32}$ for YH$_{10}$ and rare-earth hydrides [12] such as LaH$_{10}$ [12, 15, 16]. The predicted critical temperatures for LaH$_{10}$, YH$_{10}$ [12, 15, 17, 18], and CaH$_6$ [14] under pressures on the order of 0.1 $TPa$ lie within the range of $235 < T_c < 320\ K$. Although numerous theoretical predictions have been made, only a few have been confirmed through experiments. This limitation arises from the impossibility of accurately considering the anharmonicity of hydrogen atoms in calathrate-like structures within these calculations [19]. Currently, anharmonicity and quantum fluctuations can be successfully addressed to achieve more accurate predictions while maintaining a reasonable computational cost [20]. In addition to clathrate-like structures, scientists have also predicted high-temperature superconductivity in metal hydrides due to their unique crystal structures [21–23]. In these studies, the first-principles calculations are applied, including the zero-point energy considerations, in order to identify stable structures which are thermodynamically and dynamically stable [24–26]. Furthermore, several of these metals have been proposed as conventional superconductors [27].

Recently, we have verified the possibility of observing a state of electronic structural stability in a system composed of two corrugated graphene sheets with doped molecules placed at specific positions between the layers [28]. Independently, doped molecules H$_2$S, H$_3$S, and CeH$_9$ were being examined. For each doped molecule, we investigated a probable second-order phase transition with $T_c$ higher than the critical temperature of cuprate materials ($T_c = 133.5\ K$). By plotting the pressure-density digram for the valence electron of the system, we observed a second-order phase transition at $T_c = 186.3\ K$ and critical pressure ($P_c$) of $32.45\ MPa$ for $CeH_9$ doped molecules. However, no phase transition occurred for $H_2S$ and $H_3S$ at $T \geq 133.5\ K$. It remained undetermined whether the new phase indicates a superconductivity state or not.

The novelty of the present work lies in examining different distributions of $CeH_9$ doped molecules between two corrugated graphene layers to find the maximized critical temperature. Moreover, it confirms that the newly obtained phase indicates a superconducting state. Additionally, this research determines the type of superconductivity and computes characteristic properties such as, superconductivity energy gap ($\Delta$) and critical current density ($J_c$). None of these quantities were computed in our previous work [28]. The selection of graphene layers as substrates for exploring the phase transition of superconductivity is grounded in our prior researches and related studies [29, 30]. Grapheneسheet serves as a highly adsorbent substrate that can exert significant influence on thermodynamic properties of substances placed upon it. For instance, despite theoretical investigations suggesting the absence of $^3$He liquefaction in strictly two dimensions [31, 32], low-temperature heat capacity measurements conducted by Sato *et al.* [33, 34] have clearly demonstrated a liquid-gas phase transition for fluid $^3$He adsorbed on a graphite surface in two dimensions. This observed liquid state with the lowest density ever found is attributed to the exceptional adsorbent properties of graphene surfaces and their unique band structures. To further affirm of this experimental finding, we have recently verified the critical behavior of fluid $^3$He adsorbed on a graphene layer by employing the lowest-order constrained variational (LOCV) many-body method [29]. As far as we know, this has been the first theoretical research to consider the substrate deformations to verify a liquid state of fluid $^3$He in two-dimensions. The favorable agreement between our theoretical calculations [29] and experimental observations [33, 34] for two-dimensional liquid $^3$He indicates the efficacy of the LOCV method in accurately determining thermodynamic properties of strongly correlated fermions. The LOCV method has demonstrated numerous successful achievements in different branches of physics. More information regarding these accomplishments can be found in review papers [35].

This paper is outlined as follows: In Sec. II, we present the LOCV method and describe how $CeH_9$ doped molecules are distributed between the two corrugated graphene layers. In Sec. III, we explain how the thermodynamic and electrical properties of the valence electrons are computed. The final section provides a summary and conclusions.

## II. METHOD

We apply the LOCV method to compute the thermal, electrical and magnetic properties of the valence electrons in our system. This method is particularly well-suited for critical systems, as it does not introduce any free parameters into calculations [36]. Similar to Taylor expansion, the LOCV method divides the total energy ($E$) of the system into energy clusters ($E = E_1 + E_2 + ...,$). Systems with longer interactions require the computation of more energy cluster terms, which becomes progressively challenging when calculating the higher order terms. Consequently, the LOCV method is not suitable for systems with long-range interactions.

### A. Lowest-order constrained variational method

Since we intend to consider the spin-spin interactions in calculating the magnetic properties of the valence electrons, they are divided into $N^{(+)}$ spin-up and $N^{(-)}$ spin-down electrons. This division corresponds to the application of a magnetic field in the $z$-direction, as it will be described in subsection B. For an electron with the spin projection $i = +, -$, the Fermi-Dirac mean occupation number ($\rho^{(i)}$) is defined by,

$$\rho^{(i)}(k) = [e^{\beta(\varepsilon(k) - \mu^{(i)})} + 1]^{-1}, \qquad (1)$$

where $\varepsilon(k)$ is the single particle energy, $k$ is the wave vector of a valence electron, $\mu$ is the chemical potential, $\beta = \frac{1}{k_B T}$, and $k_B$ is the Boltzmann constant. The chemical potential is computed for each number density ($\mathfrak{n}$) and temperature ($T$) by the number of particles constraint as follows,

$$N = \sum_{k,i} [e^{\beta(\varepsilon(k) - \mu^{(i)})} + 1]^{-1}. \qquad (2)$$

For further investigations of magnetic properties of the system, we define the following spin order parameter ($\xi$),

$$\xi = \frac{N^{(+)} - N^{(-)}}{N}. \qquad (3)$$

The distinct property of LOCV method is the division of $N$ particles total wave function into two parts as follows,

$$\psi(1, 2, ..., N) = F(1, 2, ..., N) \Phi(1, 2, ..., N), \qquad (4)$$

where $F$ is the $N$ particles correlation function operator and $\Phi$ is the total wave function of the non-interacting particles [36]. $\Phi$ is determined according to system particles (fermion or boson). In fermionic case, the Slater determinant is applied. As it will be presented in subsection C, the interacting potentials of our system are short range; thus, we apply the Jastrow approximation as follows,

$$F(1, 2, ..., N) = \prod_{i > j} F(ij), \qquad (5)$$

where $F(ij)$ is the two-body correlation function demonstrating how particles $i$ and $j$ are correlated to each other. It should be noted that in systems with multiple interactions, the range of the interacting potential is determined by the net interaction.

In LOCV method, the total energy of the system is considered as the following cluster expansion,

$$E = \frac{\langle \Psi | \mathcal{H} | \Psi \rangle}{\langle \Psi | \Psi \rangle} = E_1 + E_2 + ..., \quad (6)$$

where $\mathcal{H}$ is the total Hamiltonian, $E_1$ is the one-body and $E_2$ is the two-body cluster energies, etc. $E_1$ and $E_2$ are defined as,

$$E_1 = \sum_i E_1^{(i)} = \sum_{k,i} \rho^{(i)}(k)\varepsilon(k), \quad (7)$$

and

$$E_2 = \frac{1}{2} \sum_{i,j,j_1,j_2} \rho^{(i)}(k_{j_1})\rho^{(j)}(k_{j_2})\langle j_1 j_2 | \omega(12) | j_1 j_2 \rangle_a, \quad (8)$$

where $j_1$ and $j_2$ are two-body state, $a$ indicates the antisymmetry of the wave function and $\omega(12)$ is the effective potential of two-body cluster defining as,

$$\omega(12) = \frac{\hbar^2}{m^*}\left[\vec{\nabla}F(\mathbf{r})\right]^2 + F^2(\mathbf{r})V(\mathbf{r}), \quad (9)$$

where $\hbar = \frac{h}{2\pi}$, $h$ is the Plank constant, $\vec{\nabla}$ is gradient operator, $V(\mathbf{r})$ is the inter-particle potential with distance $\mathbf{r}$, and $m^*$ is the effective mass of the valence electrons. To consider the spin-spin interactions, we write the two body correlation function as follows,

$$F(\mathbf{r}) = \sum_{s=0,1} f_s(\mathbf{r}) P_s, \quad (10)$$

where,

$$P_0 = \frac{1}{4}(1 - \vec{\sigma}_1 \cdot \vec{\sigma}_2), \quad (11)$$

and

$$P_1 = \frac{1}{4}(3 + \vec{\sigma}_1 \cdot \vec{\sigma}_2). \quad (12)$$

$f_0$ and $f_1$ are the spin-singlet and spin-triplet two-body correlation functions, respectively and $\sigma$ is the electron spin. By inserting Eq. (10) in Eq. (9) and doing some algebra, the spin dependent equation for $E_2$ is derived,

$$E_2 = \sum_{s=0,1} E_{2,s}, \quad (13)$$

where,

$$E_{2,s} = 2\pi\mathfrak{n} \int_0^\infty d\mathbf{r}\,\mathbf{r}^2 \left[\frac{\hbar^2}{m^*}[\nabla f_s(\mathbf{r})]^2 + f_s^2(\mathbf{r})V(\mathbf{r})\right] a_s. \quad (14)$$

In this equation, $a_0$ and $a_1$ are defined as,

$$a_0 = \frac{1}{4}\left[1 - \xi^2 + \gamma^{(+)}(\mathbf{r})\gamma^{(-)}(\mathbf{r})\right],$$
$$a_1 = \frac{1}{4}\Big[3 + \xi^2 - \gamma^{(+)}(\mathbf{r})\gamma^{(-)}(\mathbf{r})$$
$$\qquad - \left[\gamma^{(+)}(\mathbf{r})\right]^2 - \left[\gamma^{(-)}(\mathbf{r})\right]^2\Big], \quad (15)$$

where,

$$\gamma^i(\mathbf{r}) = \frac{1}{\pi^2 \mathfrak{n}} \int_0^\infty dk\, k^2 \frac{\sin(k\mathbf{r})}{k\mathbf{r}} \rho^{(i)}(k). \quad (16)$$

For more details see Ref. [36].

The two-body energy is minimized with respect to the variation of the two-body correlation function by imposing the normalization constraint [36]. This minimization is done by the following Euler-Lagrange differential equation,

$$f_s''(\mathbf{r}) + \left(\frac{2}{\mathbf{r}} + \frac{a_s'}{a_s}\right)f_s'(\mathbf{r}) - \frac{m^*}{\hbar^2}(V(\mathbf{r}) - 2\Lambda)f_s(\mathbf{r}) = 0, \quad (17)$$

where Lagrange multiplier $\Lambda$ is considered as the normalization constraint and prime symbol denotes the derivation with respect to $\mathbf{r}$. The spin dependent two-body correlation function is computed by solving Eq. (17) and then the energy clusters are obtained. The higher order clusters have lower contributions in total energy of the system; therefore, the energy clusters should be computed respectively until their contributions become ignorable in total energy calculations. Further details regarding higher orders of energy clusters are represented in Ref. [36].

### B. Magnetic susceptibility

The magnetic susceptibility ($\chi$) is a significant response function in understanding the magnetic property of materials. In our procedure, we calculate the magnetic susceptibility of the system by considering an external magnetic field in $z$-direction ($H_z$) and adding the magnetic term $-\sum_i \vec{\mu}_i \cdot \vec{H}$ to the system Hamiltonian, where the electron magnetic moment $\vec{\mu}_i$ is equal to $\vec{\mu}_i = -\frac{\mathbf{e}}{m^*}\vec{\sigma}_i$, and $\mathbf{e}$ is the electron charge. The magnetic susceptibility is obtained as follows,

$$M_z = \left(\frac{\partial \mathcal{F}}{\partial H_z}\right)_{T,n}, \quad (18)$$

$$\chi = \left(\frac{\partial H_z}{\partial M_z}\right)_{T,n}, \quad (19)$$

where, $M_z$ is the magnetization in $z$-direction and $\mathcal{F}$ is the free energy of the valence electrons.

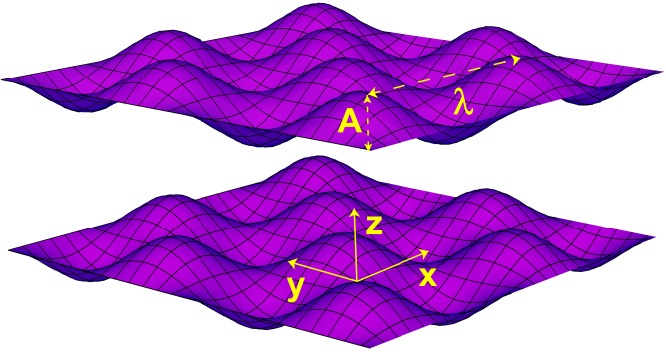

FIG. 1: A sine form corrugation is considered for the two graphene surfaces. $A$ and $\lambda$ are the amplitude and the periodicity of corrugations, respectively. The figure has no scale. The carbon atoms and honeycomb patterns are not shown for a better outlook.

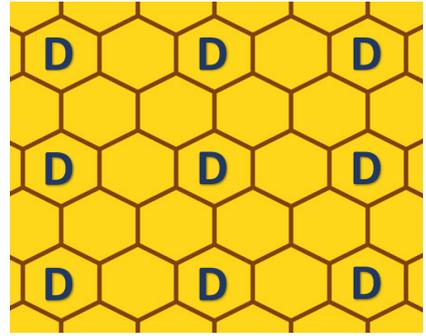

FIG. 2: A two-dimensional view of how doped molecules (D) are distributed in between the two graphene layers in x-y plane for the case $n_1 = 2$, $n_2 = 2$.

### C. System configuration

When graphene layers are applied in theoretical works, the first problem is considering the ripples of graphene surface in calculations, particularly at high temperatures. Many theoretical scientists ignore these small thermal ripples for simplicity of calculations while considering them may change the results extremely. For instance, when the ripples on the surface of graphene are disregarded in calculations, the liquid phase of a two-dimensional fluid $^3$He, placed on a graphene substrate, disappears [29]. Ripples on graphene layers are stochastic phenomena, while corrugations can be mathematically defined as periodic functions; therefore, corrugations can be considered in equations such as the Schrödinger equation. Technically, when a graphene layer is being manufactured, the manufacturer is able to consider an external corrugations on its surface to prevent the appearance of intrinsic ripples [37]. Taking into account the corrugations within graphene structure, rather than thermal ripples, has proven to be a satisfactory and accurate approach in our implemented theoretical calculations [28–30].

In present work, we consider two graphene sheets with a $C-C$ distance of 1.42 Å whose honeycomb patterns are placed exactly on top of each other, separated by a distance $l$; we call this, equilibrium distance. To introduce the deformations of the surfaces, sine-shaped corrugations with periodicity $\lambda$ and amplitude $A$ are considered for both graphene surfaces which are mathematically well-behaved functions (Fig. 1). These corrugations do not change the periodicity of the honeycomb pattern. Previously mentioned, our research focuses on exploring the electrical and thermodynamic properties of a system containing two corrugated graphene layers with intercalated CeH$_9$ doped molecules. In order to compute theses properties of the system, the coordinates of the equilibrium positions of CeH$_9$ molecules, and carbon atoms of the graphene layers are required. It should be emphasized that the equilibrium positions of atoms represent the atomic averaged positions based on their quantum probability densities. The presence of this issue poses a challenge in conducting theoretical investigations on our complicated system. To address this problem, we first establish trial positions for CeH$_9$ molecules and carbon atoms. Subsequent optimization process will refine these trial positions to determine the exact equilibrium coordinates of the system constituents. The cerium atoms of CeH$_9$ doped molecules are initially distributed periodically and precisely at the middle of the two layers of graphene at specific positions, for instance, Fig. 2 indicates a two-dimensional view of how doped molecules were distributed periodically in our previous research [28]. In this figure, the cerium atoms of CeH$_9$ molecules are initially positioned precisely at the midpoint of the line segments connecting the centers of two honeycomb structures, with one honeycomb located above and the other behind the cerium atoms. These trial positions of cerium atoms are applied for all distribution types of CeH$_9$ molecules between two graphene layers in this study.

In our earlier investigation [28], we considered just one kind of distribution of doped molecules between two graphene layers (Fig. 2), while in the present work we consider different kinds of distributions to maximize the critical temperature of the valence electrons. For this purpose, we define two unit vectors $\vec{b_1}$ and $\vec{b_2}$ for a flat graphene shown in Fig. 3, where the distance between two consequent doped molecules in horizontal and vertical directions are defined by $n_1 \vec{b_1}$ and $n_2 \vec{b_2}$, respectively with $n_1 = 2, 3, \cdots$ and $n_2 = 1, 2, \cdots$. For instance, the distribution of $n_1 = 2, n_2 = 1$ is our previous work (Fig. 2). Due to the relatively large size of CeH$_9$ molecules, the integer of $n_1 = 1$ is not permitted. The positions of carbon atoms in the sinusoidal graphene layers of Fig. 1

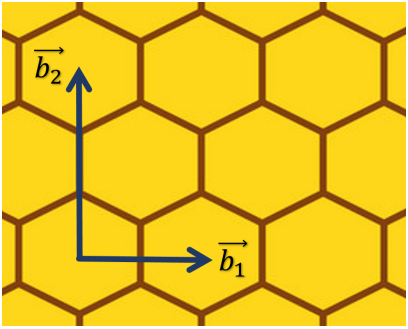

FIG. 3: Graphene honeycomb pattern. The distance between two consequent doped molecules in horizontal and vertical directions are determined by integer multiples of $\vec{b_1}$ and $\vec{b_2}$ unit vectors, respectively.

are considered as the trial positions for the carbon atoms.

In order to calculate the trial coordinates of the nine hydrogen atoms around the Ce atoms, we begin by solving the problem of how a single $CeH_9$ molecule is stabilized between two flat graphene layers. The hydrogen coordinates derived from this scenario will be considered as the trial coordinates of nine H atoms in all $CeH_9$ molecules situated between the two corrugated graphene layers. In this regard, a Ce atom is placed exactly at the midpoint of the line segment that connects the centers of the two honeycomb structures, one above and the other behind the cerium atom. We consider a hypothetical spherical space with a radius of $\approx 3$ Å centered on Ce. This sphere represents the space in which H atoms interact effectively with Ce, C, and other H atoms. The procedure for calculating this effective radius is explained in the RESULTS AND DISCUSSIONS section. The spherical space is subdivided into cubic subspaces with a volume of $10^{-3}$ Å$^3$. The free energy of a single $CeH_9$ molecule between two flat graphene layers is calculated by the LOCV method for all possible effective distributions of nine hydrogen atoms among the cubes, with each hydrogen atom placed at the center of one cube. The positions of the nine hydrogen atoms surrounding the Ce atom are determined by the configuration with the lowest free energy. The coulombic potential is considered for H-H, H-Ce, and H-C interactions, along with spin-spin interactions (Eq. (14)). The process of calculating the free energy of a single $CeH_9$ molecule between two flat graphene layers by the LOCV method mirrors the computation of the free energy of valence electrons, as elaborated in Eqs. (20), (21) and (23).

All these trial coordinates are starting points. The real equilibrium coordinates of all atoms are determined by the following variational approach. The trial positions of C, Ce, and H atoms are specified by $\vec{R}'_C$, $\vec{R}'_{Ce}$, and $\vec{R}'_H$, respectively. By using these trial positions, we are able to calculate the free energy of the lattice structure by the LOCV method. The second step involves allowing the atoms to move and find their most stable positions, which they autonomously place during the experiment.

By this procedure, we are able to adapt our theoretical calculations with experimental realities. These degrees of freedom are manipulated by adding $\delta \vec{R}'$ to atoms positions; $\vec{R}_C = \vec{R}'_C + \delta \vec{R}'_C$, $\vec{R}_{Ce} = \vec{R}'_{Ce} + \delta \vec{R}'_{Ce}$, and $\vec{R}_H = \vec{R}'_H + \delta \vec{R}'_H$. The free energy of the lattice structure is minimized with respect to variation of $\delta \vec{R}'_C$, $\delta \vec{R}'_{Ce}$, and $\delta \vec{R}'_H$ to find the optimized positions of all atoms; therefore, $\vec{R}$ vectors represent the real equilibrium positions of atoms in the lattice structure.

The electrons of both carbon atoms and doped molecules can be classified into valence (e) and non-valence electrons. The valence electrons are exactly the thermodynamic system we intend to investigate its critical behavior. In our methodology, these electrons are not attributed to any specific atom. They flow in crystal lattice like a fluid to make conductivity or superconductivity; therefore, we call them fluid electrons. Meanwhile, the non-valence electrons are bound to nuclei causing shielding effect. These bindings prevent their involvement in conductivity or superconductivity. In our calculations, all carbon, cerium, and hydrogen atoms are considered as ions with effective charge $q$ where fluid electrons move between them. These ions are $Ce^{9+}$, $C^{4+}$, and $H^+$ with no fluid electrons. The effective charge $q$ of each ion is determined by its positive nucleus and shielding effect resulting from non-valence electrons.

In LOCV method, the single particle energy state $\varepsilon(k)$, the single particle wave function $\Phi(k)$, and the interacting potentials are essential input parameters. By computing these components, all thermodynamic functions are obtained. To compute $\varepsilon(k)$ and $\Phi(k)$, the following Schrödinger equation is solved for a fluid electron positioned at $\vec{\mathfrak{r}} = x\hat{i} + y\hat{j} + z\hat{k}$ interacting with the electric potential generated by the carbon and doped ions,

$$-\frac{\hbar^2 \nabla^2}{2m^*}\Phi + U\Phi = \varepsilon \Phi, \qquad (20)$$

where,

$$U = \sum_{i=1}^{N_1} V_{e-C}(\vec{R}_{C_i}, \vec{\mathfrak{r}}) + \sum_{j=1}^{N_2} V_{e-D}(\vec{R}_{D_j}, \vec{\mathfrak{r}}). \qquad (21)$$

$\nabla^2$ is the Laplacian operator, and $N_1 = 54$ and $N_2$ are the number of carbon and doped ions, respectively that have effective interactions with a fluid electron. These effective numbers are calculated by evaluating the distances at which the e-C and e-D interacting potentials decrease by two orders of magnitude from their maximum absorbing absolute values.

A fluid electron feels three distinct potentials, (1) electron with carbon ions of graphene lattice (e-C), (2) electron with doped ions (e-D), and (3) electron with other valence electrons (e-e). The interactions e-C and e-D are considered in Eq. (21), while e-e is considered in Eq. (8) and higher energy clusters. The general solution of Eq.

(20) is as follows,

$$\Phi(x, y, z, \vec{k}) = \mathcal{A}.u_1(x, y, k_x, k_y)u_2(z, k_z).e^{(ik_x x + ik_y y)}, \quad (22)$$

where $\mathcal{A}$ is the normalization constant. The consideration of this wave function arises from the periodicities of both the graphene corrugations and the distribution of $CeH_9$ molecules. Due to these periodicities, the wave function is written in Bloch [38] form where $u_1(x, y, k_x, k_y)$ satisfies the Born-Von Karman periodic boundary condition within x-y plain. The dependency of the wave function on $z$ and $k_z$ is considered by the term $u_2(z, k_z)$. By inserting the boundary conditions into Eq. (20), the Schrödinger equation is solved by shooting method, thereby $\varepsilon(k)$ and $\Phi(k)$ are obtained. The total energy of the fluid electrons is then calculated by inserting $\varepsilon(k)$ and $\Phi(k)$ into energy clusters and adding them together. It is more convenient to work with free energy ($\mathcal{F} = E - TS$) instead of total energy where $S$ is the entropy of the system,

$$S(m^*, T) = -k_B \sum_{k,i} \left[ \left(1 - \rho^{(i)}(k)\right) \ln\left(1 - \rho^{(i)}(k)\right) + \rho^{(i)}(k) \ln\left(\rho^{(i)}(k)\right) \right]. \quad (23)$$

The effective mass is treated as a variational parameter and the free energy of the fluid electrons is minimized with respect to the variation of $m^*$. By this procedure, the e-phonon interactions are considered in the free energy calculations. These interactions play a crucial role in determining the critical behavior of the systems. In the next section, it will be demonstrated that by disregarding these interactions, the critical behavior of the fluid electrons disappears.

### III. RESULTS AND DISCUSSIONS

The LOCV approach is widely recognized as a highly effective many-body technique in diverse fields of scientific researches. It has been successfully employed to verify numerous experimental findings, providing valuable insights into various phenomena [35]. Before presenting the results of our system, we have applied the LOCV method to reanalyze a portion of the experimental survey conducted by Chen et al. [39] to evaluate the accuracy and reliability of our methodology. Our aim is to compare our recalculated results with those obtained through empirical observation to assess the validity of our approach against the real-world data. These calculations are not related to our main purpose of verifying the critical behavior of intercalated $CeH_9$ molecules between graphene layers. Chen et al. experimentally and theoretically discovered the superconductivity state in two new phases, $Fm\bar{3}m$-$CeH_{10}$ and $P6_3/mmc$-$CeH_9$. The experimental results of Ref.[39] and our theoretical results (LOCV method) for superconductivity in cerium superhydride ($P6_3/mmc$-$CeH_9$) are presented in table I. $P_{e-c}$ is the external critical pressure applied on the system. This table illustrates a satisfactory agreement between the LOCV results and experimental data, with a difference of approximately $\simeq 7\%$, while the results obtained from Migdal-Eliashberg equations and the Allen-Dynes formula [40] indicate differences exceeding 40% with respect to experimental data. For more information about the results of Migdal-Eliashberg equations and Allen-Dynes formula see Ref. [39].

Now let us refocus on our primary research objective. First, we aim to clarify how the number of hydrogen ($\eta$) atoms is determined in intercalated $CeH_\eta$ molecules into the graphene layers. For this purpose, the free energies per particle of lattice structure for $\eta$ values of $4, 5, 6, 7, 8, 9,$ and $10$ are presented in table II. The resulting data are calculated by the LOCV method with $l = 8.09$ Å for the distribution $n_1 = 11$, $n_2 = 6$ at $T = 198.61$ K. It will be seen later that these values for $l$, $T$, and the distribution type of cerium superhydrides are the conditions associated with the highest $T_c$ superconductivity discovered in this study. Interestingly, among the different values of $\eta$, $CeH_9$ exhibits the lowest free energy indicating a more stable state compared to the others, particularly $CeH_4$, which may seem more compatible with the electronic configuration of cerium atom at first glance. Based on these free energy calculations, $CeH_9$ superhydrides are considered as doped molecules between the two corrugated graphene layers in the current study.

In the METHOD section, we discussed the process of determining the positions of nine hydrogen atoms in a single $CeH_9$ molecule intercalated between two flat graphene layers. These positions are presented in spherical coordinates in table III, where $\theta$ and $\phi$ represent the polar and azimuthal angles, respectively. These results are calculated according to the directions of the axes in the coordinate system illustrated in Fig. 1. The origin of the coordinate system is set at the center of the Ce atom. Fig. 4 provides a partial 3-dimensional view of the stabilization of a $CeH_9$ molecule between two graphene layers, with an error margin of 0.05 Å in each direction. We were able to compute these quantities more precisely. However, this level of precision is unnecessary since they serve as the trial coordinates in the process of determining the accurate equilibrium positions of the structure. The free energy results show a notable increase for H-Ce distances exceeding approximately 3 Å. Consequently, we constrain the free energy minimization process to distances less than 3 Å.

The critical temperature of the fluid electrons are calculated for all cases of $n_1 = 2, 3, \cdots, 100$ and $n_2 = 1, 2, \cdots, 100$, resulting in a total of 9900 distinct distributions of $CeH_9$ molecules. For each distribution of $CeH_9$ molecules, the total energy per electron are computed for the fluid electrons by adding the energy clusters upto $E_3$ term. In this study, all values of the energy clusters are presented on a per-particle basis. Due to the negligible contributions of the clusters higher than three-body energy ($E_4, E_5, \cdots \lesssim 10^{-3} \ Kk_B$), these terms are ignored. The entropy of the fluid electrons is calculated by Eq. (23), and the corresponding free energy is minimized with respect to $m^*$, $A$, $\lambda$, and $l$ across all temperatures and densities. This process is known as structural relaxation.

| Methodology | $T_c$ (K) | $P_{e-c}$ (GPa) | $J_c$ (A/cm$^2$) |
|---|---|---|---|
| Experiment | $\simeq 102$ | $\simeq 130$ | $\sim 10^8$ |
| LOCV | 109.51 | 138.97 | $\sim 10^8$ |

TABLE I: The experimental results for superconductivity in cerium superhydride ($P6_3/mmc$-CeH$_9$) [39] compared to theoretical calculations obtained by the LOCV method.

| $\eta$ | 4 | 5 | 6 | 7 | 8 | 9 | 10 |
|---|---|---|---|---|---|---|---|
| Free energy per particle (MeV) | 1.782 | 2.763 | 2.867 | 2.902 | 2.914 | 1.771 | 1.813 |

TABLE II: The free energy per particle of the lattice structure for $\eta$ values of $4, 5, 6, 7, 8, 9$, and $10$ at $T = 198.61 \ K$. The most stable lattice structure is related to $\eta = 9$ (CeH$_9$) with the minimum free energy per particle.

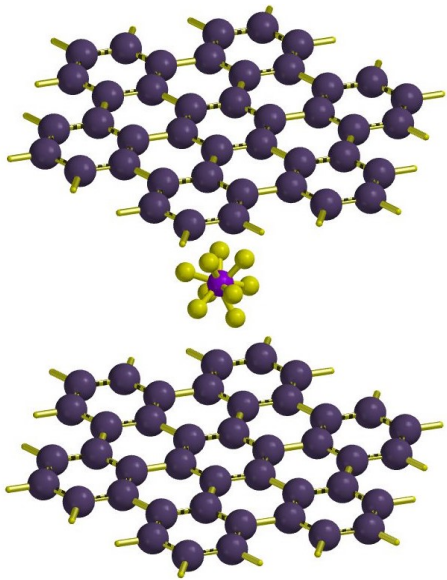

FIG. 4: Three-dimensional representation of a single CeH$_9$ doped molecule positioned between two graphene layers. For clarity, the atom sizes are adjusted, the interlayer distance is increased, and the H-Ce atomic distances are slightly reduced proportionally in this figure. The precise polar and azimuthal angles of the nine hydrogen atoms are accurately depicted.

### A. Second-order phase transition of fluid electrons

In systems with phase transition, free energy diagram is a valuable tool for identifying transition type and computing critical point properties. To illustrate this procedure, the free energy per electron is presented versus $\mathfrak{n}^{-1}$ for distribution $n_1 = 11$, $n_2 = 6$ at $T = 197 \ K$ in Fig. 5. Due to the presence of a non-stable zone around $\mathfrak{n}^{-1} = 24 \ \text{Å}^3$, this diagram effectively illustrates a first-order phase transition for the fluid electrons and also highlights a fundamental aspect of superconductivity, spontaneous symmetry breaking. A transition from a normal state to a structural stability state. It will be justified later that this transition indicates a superconductivity phase transition. In order to determine the boundaries of the transition process, we apply the double-tangent Maxwell construction which is depicted by the dashed line in Fig. 5. Further details about this construction can be found in Ref. [41]. The transition process initiates at $\mathfrak{n}^{-1} = 24.22 \ \text{Å}^3$ (point A), where all fluid electrons are in the normal state. As the density increases, they gradually undergo a transition towards the state of structural stability. Finally, at $\mathfrak{n}^{-1} = 23.69 \ \text{Å}^3$ (point B), all fluid electrons enter the new phase and the transition process is terminated. To find the critical point, it is necessary to increase the temperature. As the temperature increases, the length of the dashed line decreases since the points A and B move closer together. Eventually, at a characteristic temperature ($T_c = 198.61 \ K$), these two points converge into a single point. This distinctive point represents the critical point, which signifies a second-order phase transition for the fluid electrons.

While some theoretical scientists apply the Allen-Dynes equation to calculate the $T_c$ of superconductivity, we employ the Maxwell construction method as an alternative approach. The Allen-Dynes equation has been derived from Eliashberg theory with the approximation of ignoring the momentum dependency of the Eliashberg function, whereas the Maxwell construction approach is able to be applied at all temperatures with no limitations if the free energy of the system is available. An outstanding approach employed in this work involves considering one wave function for both $T > T_c$ and $T < T_c$ cases, without changing the Hamiltonian form in these cases; therefore, the phase transition and spontaneous symmetry breaking occur intrinsically by the collective behavior of the fluid electrons. By maintaining consistency in the wave function across the critical point, this methodology provides a comprehensive understanding of how the fluid

| atom | radius (Å) | $\theta$ (degree) | $\phi$ (degree) | atom | radius (Å) | $\theta$ (degree) | $\phi$ (degree) | atom | radius (Å) | $\theta$ (degree) | $\phi$ (degree) |
|---|---|---|---|---|---|---|---|---|---|---|---|
| Ce | 0 | - | - | H | 2.09 | 76.07 | 120.40 | H | 2.11 | 116.01 | 194.26 |
| H | 2.10 | 95.65 | 44.38 | H | 2.10 | 147.31 | 111.52 | H | 2.08 | 142.41 | 317.17 |
| H | 2.03 | 20.43 | 22.23 | H | 2.10 | 52.53 | 191.24 | | | | |
| H | 1.98 | 86.76 | 258.57 | H | 2.13 | 73.92 | 328.53 | | | | |

TABLE III: Optimized equilibrium coordinates of a single $CeH_9$ molecule intercalated between two flat graphene layers in spherical coordinates. The results are computed according to the directions of the axes in the coordinate system depicted in Fig. 1. The origin of the coordinate system is placed at the cerium atom.

| Critical parameter | distribution (1) | distribution (2) | distribution (3) | distribution (4) |
|---|---|---|---|---|
| $T_c$ (K) | 198.61 | 198.52 | 198.45 | 186.3 |
| $\mathfrak{n}_c$ ($nm^{-3}$) | 43.32 | 43.14 | 42.72 | 41.3 |
| $P_c$ ($MPa$) | 34.80 | 34.66 | 34.25 | 32.45 |

TABLE IV: Among all 9900 different distributions of $CeH_9$ molecules, considered separately between the two graphene layers, three distributions with $n_1 = 11, 10, 9$ and $n_2 = 6, 7, 7$ have the highest $T_c$ labeling as distributions 1, 2, and 3, respectively. Distribution 4 with $n_1 = 2$, $n_2 = 1$ is our previous work.

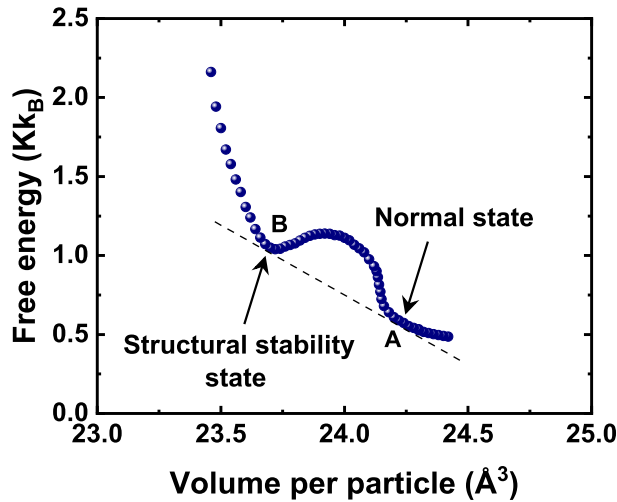

FIG. 5: The free energy per electron as a function of volume per particle at $T = 197$ K for distribution $n_1 = 11$, $n_2 = 6$. The dashed line shows the double-tangent Maxwell construction for finding the starting and ending points of the first-order phase transition. At point B, all fluid electrons are in structural stability state, while at point A, they are in a normal state.

electrons undergo the phase transition.

The free energies and related critical points of the fluid electrons are computed for all 9900 different distributions of $CeH_9$ molecules, separately. The results of our previous work and the first three distributions with the highest critical temperatures are shown in table IV. In this table, distributions $1, 2, 3$, and $4$ correspond to the values of $n_1 = 11, 10, 9, 2$ and $n_2 = 6, 7, 7, 1$, respectively. Distribution (4) is our previous work. Among 9900 verified cases, the case $n_1 = 11$, $n_2 = 6$ indicates the highest $T_c$ which is approximately 65 K greater than the $T_c$ of the cuprate superconductor. It should be highlighted that the results of megapascal pressures listed in table IV are not external pressures applied to the outer surfaces of the two graphene layers. Instead, they represent the interior pressures between the fluid electrons, arising from the fermionic nature of electrons and their interactions with the environment. Meanwhile, the other fabricated superconductors with $T_c > 133.5$ K require extremely higher external pressures on the order of $GPa$ [1, 42]. The comparison between the results of superconductivity in $P6_3/mmc$-$CeH_9$ (table I) and intercalated $CeH_9$ doped molecules between the two corrugated graphene layers is fascinating. While the first system necessitates an external pressure of 138.97 $GPa$, no external pressure needs to be applied on the exterior surfaces of the graphene layers in the second system. Despite this difference, there is a remarkable increase in critical temperature from 109.51 K to 198.61 K. These findings strongly indicate the crucial role played by graphene layers in eliminating the requirement of external pressures and enhancing the critical temperature. It is important to note that the internal pressure within the fluid electrons remains constant over time, which does not lead to the expansion of the structure. This constancy is maintained as the internal pressure of the fluid electrons is calculated at a thermodynamic equilibrium state after establishing the stabilized ionic structure through the minimization of lattice free energy. All thermodynamic quantities of the system are computed after structural relaxation.

Based on our calculations, the effective mass of fluid electrons indicates a variation range of $0.19 \leq \frac{m^*}{\mathfrak{m}} \leq 0.46$, where $\mathfrak{m}$ represents the free electron mass. These results are related to distribution $n_1 = 11$, $n_2 = 6$, near the critical point at $198.1 \leq T \leq 199.1$ K. The lower values of $m^*$ compared to $\mathfrak{m}$ indicate an absorbing interaction among fluid electrons. This interaction facilitates the formation of Cooper pairs and signifies the importance of e-phonon couplings in occurring the second-order phase transition. These results are the consequent of employing $m^*$ instead of $\mathfrak{m}$, and minimizing the free energy with respect to $m^*$. This replacement enables us to consider the e-phonon interactions into free energy calculations. Three primary factors contribute to observing the critical behavior in fluid electrons are the band structure of the corrugated graphene, the numerous hydrogen atoms in $CeH_9$ superhydrides, and considering $m^*$ as a variational parameter.

This is evident as no phase transition occurs by replacing $m^*$ with $\mathfrak{m}$ or applying hydrides with fewer hydrogen atoms, such as $CeH_4$ doped molecules, at $T \geq 133.5\ K$. The optimized values of $l$, $A$, and $\lambda$ for distribution (1) are computed by minimizing the free energy as follows: $l = 8.09$ Å, $A = 5.76$ Å, and $\lambda = 27.2$ Å; all calculations for distribution (1) are performed by considering these values for related variational parameters. The critical temperature indicates a strong dependence on variational parameters such that the observation of phase transition for distribution (1) is limited to $7.14 \leq l \leq 8.46$ Å, $4.31 \leq A \leq 6.88$ Å, and $25.23 \leq \lambda \leq 29.06$ Å for temperatures greater than $133.5\ K$. It is surprising that among all 9900 verified distributions, the maximum $T_c$ belongs to the case whose distribution periodicity of $CeH_9$ molecules is the closest to the periodicity of the graphene corrugations. The underlying reason behind this observation remains unclear to us. Nevertheless, it presents a trial distribution that can serve as a promising starting point for further investigations.

This new phase of structural stability in fluid electrons is occurred due to the hydrogen- and carbon-abundant materials that can provide phonon spectra essential in formation of Cooper pairs. [43, 44]. Including the variational parameter $\delta \vec{R}'$ in our calculations is another important reason in observing the phase transition. Ignoring this parameter in free energy calculations results in the disappearance of the second-order phase transition. The range of variations for $\delta \vec{R}'_C$, $\delta \vec{R}'_{Ce}$, and $\delta \vec{R}'_H$, are $1.57 \leq \delta \vec{R}'_C \leq 1.73\ pm$, $0.11 \leq \delta \vec{R}'_{Ce} \leq 0.23$ Å, and $0.24 \leq \delta \vec{R}'_H \leq 0.92$ Å, respectively, for distribution $n_1 = 11$, $n_2 = 6$. Based on these results, as we expected, the carbon ions are strictly bounded to the graphene lattices. Due to the heavier mass of Ce ions and their surrounding by nine hydrogen ions, they tend to maintain their primary positions, while H ions demonstrate a greater mobility owing to their lighter masses and high chemical reactivity. Theses results reveal that when $CeH_9$ molecules are placed at specific sites between the graphene layers, the primary positions of C, Ce, and H ions do not change largely. Importantly, our methodology accounts for these movements by minimizing the free energy of the ionic structure. The constrained mobilities of C, Ce, and H ions are significant as they provide com-pelling evidence that this theoretical investigation can be verified through experimentation. If the ionic movements were larger, they would change the boundary conditions for solving the Schrödinger equations and significantly complicate the calculations.

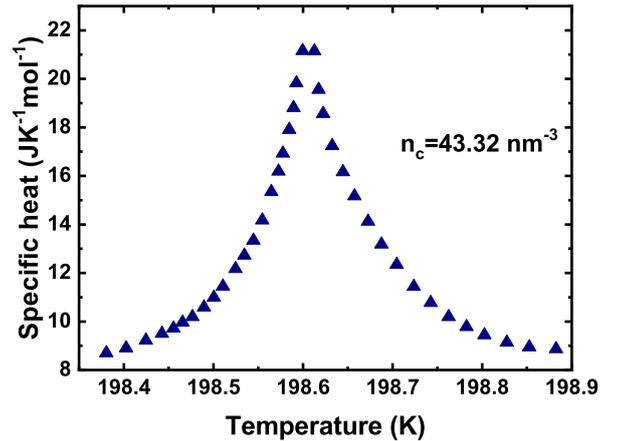

FIG. 6: The specific heat of fluid electrons at $\mathfrak{n}_c = 43.32\ nm^{-3}$ for distribution (1). Due to the second-order phase transition, the specific heat diverges by approaching the critical point.

The divergence behavior of response functions is a characteristic property of second-order phase transitions. In order to illustrate this critical behavior, the specific heat at constant volume ($C_v$) for fluid electrons is demonstrated in Fig. 6 for distribution $n_1 = 11$, $n_2 = 6$. Near the critical point, the fluctuations of the free energy are not ignorable since they enlarge as the system approaches the critical point. Consequently, the short-range interactions of e-e and e-phonon give rise to long-range correlations among the fluid electrons. This is precisely why $C_v$ indicates scaleless behavior in the vicinity of the critical point. The magnetic susceptibility ($\chi$) of the fluid electrons is plotted in Fig. 7 as another response function for distribution $n_1 = 11$, $n_2 = 6$ near the critical point. These results are obtained by considering an external magnetic field along the $z$-direction and calculating the system's magnetic response. The magnetic susceptibility behaves normally (very small and positive) at $T > T_c$ while it reduces suddenly into $\simeq -1$ just below the critical temperature; for instance, $\chi = -0.91$ at $T = 198.4\ K$. This is another sign of the second-order phase transition from the normal state (paramagnetism) to the structural stability state (almost perfect diamagnetism). Although the Meissner effect is associated with $\chi = -1$, the diamagnetic part shown in Fig. 7 demonstrates the Meissner effect. Because the accuracy of approximations used in computing the total wave function can influence the values of $\chi$. Moreover, it will be shown that the phase transition signifies a superconductivity state.

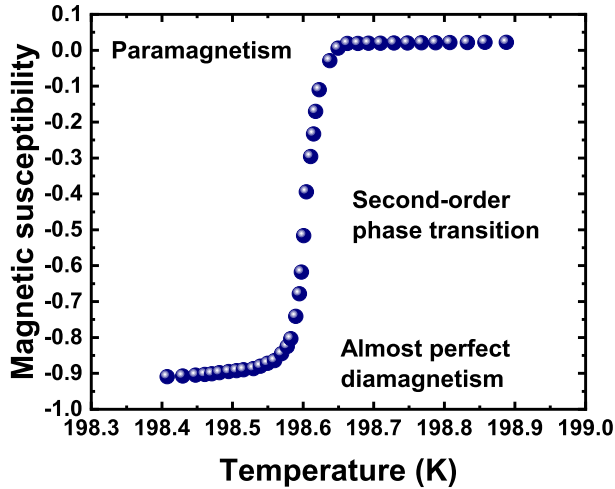

FIG. 7: The Magnetic susceptibility of the fluid electrons at $\mathfrak{n}_c = 43.32 \ nm^{-3}$ for distribution (1) of CeH$_9$ doped molecules. Above $T_c = 198.61 \ K$, $\chi$ is positive and trivial while at $T \lesssim T_c$, the magnetic susceptibility rapidly decreases to $\simeq -1$, signifying an almost perfect diamagnetism.

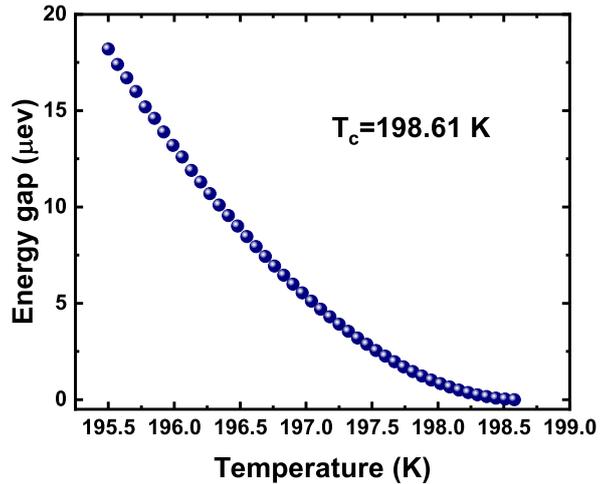

FIG. 8: The energy gap of superconductivity, near the critical point, for distribution (1).

### B. Determining the type of superconductivity and its properties

According to $\chi \simeq -1$, what is now important is identifying this new phase of the fluid electrons, where large orbital diamagnetism or superconductivity seem probable. However, the notion of large orbital diamagnetism is disregarded due to the significantly small effective mass of the valence electrons in these materials [46, 47]. Conversely, in our system, $m^*$ is not relatively small. Furthermore, in large orbital diamagnetism, the magnitude of Landau diamagnetism is inversely proportional to effective mass [48], whereas this proportion is not observed in our results. Due to the observation of the second-order phase transition and almost full diamagnetism, superconductivity is the most probable case. Second-order phase transition is a necessary condition for superconductivity while the full diamagnetism is the sufficient one. If $\chi$ was exactly $-1$ in our system, we could confidently conclude that the new phase indicates a superconductivity state; therefore, it is important to investigate the energy gap of the fluid electrons to determine whether this phase transition corresponds to a state of superconductivity or not. The superconductivity energy gap is calculated by the total energy (Eq. (6)) per particle for fluid electrons and plotted in Fig. 8 at $T < T_c$ for distribution (1). The temperature dependency of $\Delta$ is completely compatible with the following superconductivity energy gap relation [49],

$$\Delta \sim \left(1 - \frac{T}{T_c}\right)^j \left(1 + b\frac{T}{T_c}\right), \qquad (24)$$

where $j$ is a critical exponent and $b$ is a constant. $j$ and $b$ are 1.86 and 1.32 for fluid electrons of distribution (1), respectively.

The calculation of the superconductivity energy gap not only confirms that the new phase indicates a superconductivity state but also enables the determination of its type. The characterization of conventional and unconventional superconductors is a fundamental aspect in the field of condensed matter physics. Distinguishing between these two classes of materials is crucial as it allows for an in-depth investigation into their distinct properties and provides insights into the underlying physical mechanisms governing superconductivity. Calculating the reduced gap ratio, $\alpha = \frac{2\Delta_0}{k_B T_c}$, is an effective approach to determine the type of superconductors, where $\Delta_0$ is the superconducting energy gap at absolute zero temperature. For conventional superconductors $\alpha \simeq 3.5$ while it is completely different for unconventional superconductors (usually greater than 3.5) [50]. For fluid electrons of distribution (1), this gap ratio is obtained to be 7.21, which is nearly twice the value predicted by the BCS theory in the weak-coupling regime. When the gap ratio exceeds the BCS threshold, it signifies that the superconductor is categorized as a high-coupling system. Consequently, employing the Eliashberg formalism becomes highly suitable for analyzing the system behavior within this particular regime. The gap ratio $\alpha = 7.21$ indicates that the intercalated CeH$_9$ doped molecules between the two graphene layers is a new unconventional superconductor, representing fundamental similarity with Hund's superconductors [51]. In these superconductors, although $T_c$ and $\Delta_0$ are material dependent, the ratio $\alpha$ is a material-independent universal constant $\sim 7.2$ [52]. This universal aspect of superconductivity is found by Lee et al. originating from a Hund's metal state at high-temperatures.

To date, substantial efforts have been dedicated to improve the fabrication of superconductor wires and tapes, with a primary focus on increasing the critical current density, since $J_c$ plays an important role in determining the efficiency of superconductors in technology. The significance of this efficiency is such that there is a preference for employing lower $T_c$ superconductors to achieve appropriate $J_c$ values. The $J_c$ of cuprate wires, $Bi_2Sr_2CaCu_2O_8$ and $Bi_2Sr_2Ca_2Cu_3O_{10}$, with the powder-in-tube technology, is on the order of $\sim 10^6 \; A/cm^2$, and for the rare-earth barium copper oxide with the technology of coated conductors is $\sim 10^7 \; A/cm^2$ [53]. These materials are widely recognized as high-temperature superconductors, known for their high critical current density. In our system, the critical current density of the fluid electrons is calculated based on $\Delta_0$. For the case of $n_1 = 11$, $n_2 = 6$, $J_c$ is determined to be $\sim 10^7 \; A/cm^2$. This high critical current density with $T_c \simeq 199 \; K$ makes the system of the two graphene layers containing $CeH_9$ molecules a suitable material for fabricating high-$T_c$ superconducting wires.

## IV. SUMMARY AND CONCLUSIONS

In the present study, we have verified a superconductivity phase transition in valence electrons of a system consisting of two corrugated graphene sheets with intercalated $CeH_9$ doped molecules. For this purpose, we have examined 9900 distinct distributions of $CeH_9$ molecules and computed their corresponding $T_c$s. Among the considered distributions, the highest critical temperature is related to distribution (1) in table IV with $T_c = 198.61 \; K$ which is about $65 \; K$ greater than the critical temperature of the cuprate superconductor $HgBa_2Ca_2Cu_3O_{8+\delta}$. It should be emphasized that no external pressure has been applied on the outer surfaces of the graphene sheets while the internal critical pressure among the valence electrons (for instance, $P_c = 34.80 \; MPa$ for distribution (1)) is associated with the interactions of them with the environment and the fermionic total wave function.

We also have calculated the specific heat and magnetic susceptibility of the valence electrons in adjacency of the critical point. The power law behavior of the specific heat demonstrates a second-order phase transition. Furthermore, at $T \lesssim T_c$, the magnetic susceptibility experiences a sudden decrease towards $\simeq -1$. According to this almost full diamagnetism and the second-order phase transition, we expected that this structural stability of the valence electrons indicates a superconductivity state. To make it clear whether the transition is related to superconductivity or not, we have computed the energy gap of the valence electrons. The temperature-dependent energy gap calculations for the valence electrons confirms that the new phase exhibits a superconductivity state and classifies the system as an unconventional superconductor. The main reasons for observing this phase transition are the abundance of hydrogen atoms in $CeH_9$ doped molecules and the presence of carbonaceous layers, which create strong electron-phonon couplings necessary for the formation of Cooper pairs. The strong couplings of this unconventional superconductivity result in achieving high critical current density on the order of $\sim 10^7 \; A/cm^2$ which makes it suitable for manufacturing high-temperature superconductor wires.

In the preceding paragraph, we outlined the aims of our present work. However, during the computation of thermodynamic properties of the system, we encountered an interesting and unexpected relationship between two variational parameters which we are currently unable to justify. According to LOCV method, the free energy of the system is minimized with respect to some variational parameters. One of these parameters in the present work is the periodicity of surface corrugations of the two graphene layers ($\lambda$). Based on our results, the maximum $T_c$ is related to the case that $\lambda$ is closest to the distribution periodicity of $CeH_9$ doped molecules. In other words, the critical temperature decreases as $\lambda$ and the distribution periodicity move away from each other. The reason behind this correlation remains unknown for us. However, it can be applied in further researches in superconductivity of corrugated graphene layers containing doped molecules to achieve higher critical temperatures. It would be also interesting to consider other transition metals instead of cerium for increasing the critical temperature in future investigations. These metals are good candidates for the formation of superhydrides at high temperatures, making them highly suitable for finding high-temperature superconductors.

---